\documentclass{IEEEtran}
\usepackage[pdftex]{graphicx}
\usepackage{color,soul}
\usepackage[table,xcdraw]{xcolor}
\usepackage{amsmath}
\usepackage{cite}
\usepackage{url}
\usepackage{upgreek}
\usepackage{placeins}
\usepackage{subcaption}
\usepackage{amssymb}
\usepackage{amsfonts}
\usepackage{textcomp,nicefrac}
\usepackage{gensymb}

\def\BibTeX{{\rm B\kern-.05em{\sc i\kern-.025em b}\kern-.08em
T\kern-.1667em\lower.7ex\hbox{E}\kern-.125emX}}
\begin{document}
\bstctlcite{IEEEexample:BSTcontrol}
\title{FPGA-Based RoCEv2-RDMA Readout Electronics for the CTAO-LST Advanced Camera}
\author{F.~Marini, M.~Bellato, A.~Bergnoli, D.~Corti, A.~Griggio, R.~Isocrate, L.~Modenese, M.~Toffano, C.~Arcaro, F.~Di Pierro, M.~Mariotti, M.~Mi, P.~Wang
\thanks{This work was supported by the European Union - Next Generation EU, Mission 4 Component 1 CUP C53C22000430006}
\thanks{F.~Marini, M.~Bellato, A.~Bergnoli, A.~Griggio, R.~Isocrate, L.~Modenese, M.~Toffano and C.~Arcaro are with INFN Section of Padova, Padua, Italy. (e-mail: filippo.marini@pd.infn.it)
}
\thanks{M.~Mariotti is with University of Padova, Padua, Italy and INFN Section of Padova, Padua, Italy}
\thanks{F.~Di Pierro is with INFN Section of Torino, Turin, Italy}
\thanks{D.~Corti was with INFN Section of Padova, Padua, Italy and is now with INFN-TIFPA, Trento, Italy.}
\thanks{M.~Mi and P.~Wang are with DatenLord Technology Co., Ltd.}
\thanks{\newline © 2025 IEEE. Personal use of this material is permitted. 
Permission from IEEE must be obtained for all other uses, in any current or future media, 
including reprinting/republishing this material for advertising or promotional purposes, 
creating new collective works, for resale or redistribution to servers or lists, 
or reuse of any copyrighted component of this work in other works.}
\thanks{\newline Accepted for publication in IEEE Transactions on Nuclear Science. DOI: 10.1109/TNS.2025.3599615}
}
\maketitle

\begin{abstract}
CTAO’s (Cherenkov Telescope Array Observatory) largest telescopes type, the LST (Large-Sized Telescope), are being installed at the northern site of the Cherenkov Telescope Array (CTA) at the Observatorio del Roque de los Muchachos on the Canary island of La Palma. Their aim is to capture the lowest-energy gamma rays of the observatory. The hereby proposed readout electronics architecture, serving as a proof-of-concept for its advanced camera upgrade, relies on a custom high-channel count fast sampling hardware digitizer board acting as a Front-End. The design includes a versatile pre-amplification stage and high-speed serial links for streaming JESD204C-compliant data at rates approaching 12 Gb/s per lane.
The data get transferred to Back-End electronics for a first data-processing and trigger before being transmitted to event-building servers through 10 Gb/s Ethernet links. The performance of the link is exploited by implementing RDMA communication in hardware, thanks to a RoCEv2 core written in Bluespec SystemVerilog, enabling the possibility of transfer data directly to processing units without CPU intervention. Hardware design and characterization of the Front End board are reported, as well as a detailed description and tests of the Back End RDMA firmware.
\end{abstract}


\section{Introduction}
\label{sec:introduction}

\begin{figure}[t]
\centering
\includegraphics[width=3.1in]{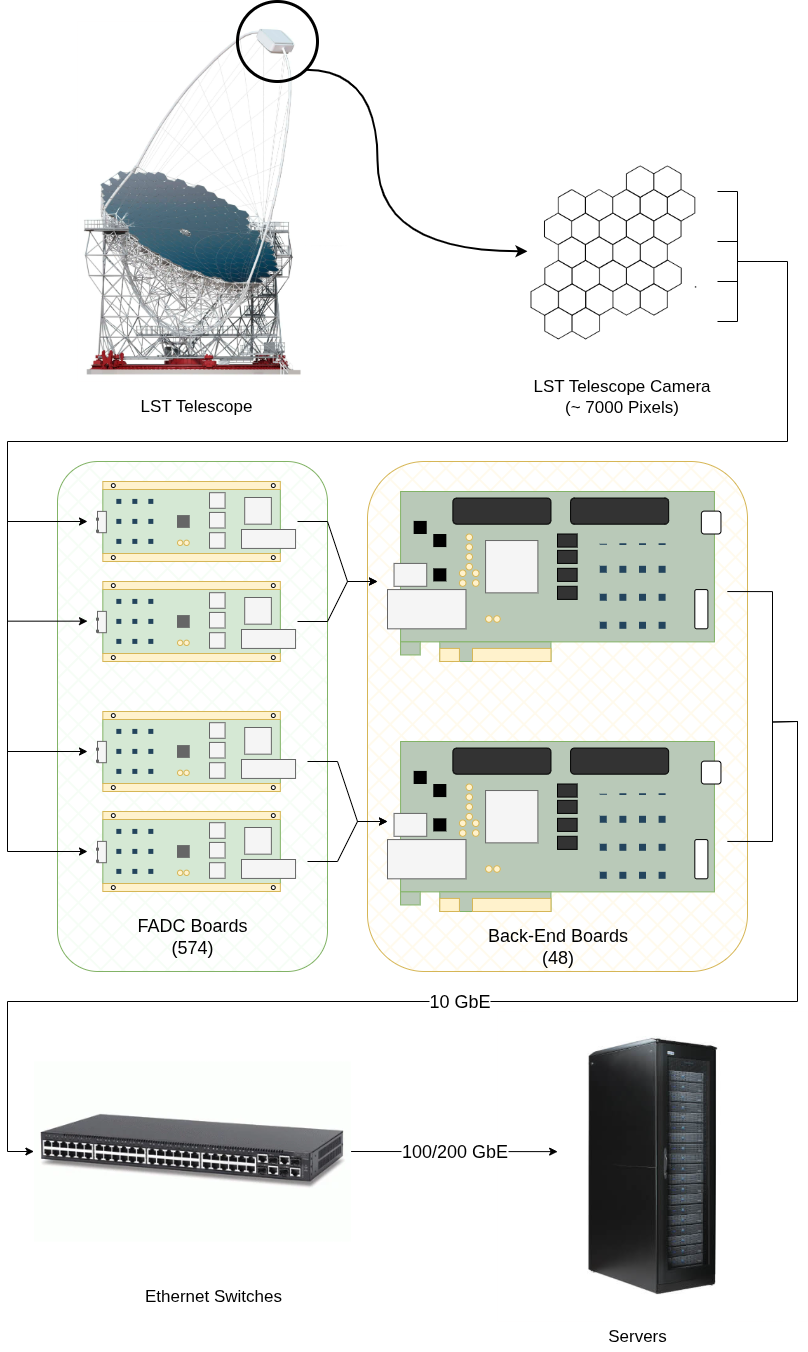}
\caption{Block diagram for the overall final system.}
\label{sketch}
\end{figure}

\IEEEPARstart{T}{he} Cherenkov Telescope Array Observatory (CTAO) is considering the adoption of Silicon Photo Multipliers (SiPMs) for their Large Size Telescopes (LST) array, specifically for a new design known as the Advanced Camera, later referred as AdvCam. The implementation of this technology poses some practical challenges, such as a higher night sky background rate due to higher sensitivity in the infra-red and longer signals compared to legacy PhotoMultiplier Tubes (PMTs). Moreover, the camera adopts smaller pixels' size, increasing the number of channels by a factor of 4, to approximately 7000 pixels. In return, the use of SiPMs enables a higher duty cycle, improved robustness, and better angular resolution thanks to finer image granularity compared to the PMT option \cite{intro}. The higher background rate (in the order of GHz) combined with the larger image granularity requires a factor of $\sim$10x increase in data throughput if a legacy readout approach is used \cite{lst_readout}. A way to tackle these issues is by designing a fully digital readout able to continuously acquire data from all sensors, so that real-time hardware-based event selection algorithms can be used as close as possible to the sensors to perform trigger decisions and increase rejection ratio, thereby reducing data bandwidth requirements. The planned AdvCam design foresees a total of 574 readout boards, required to instrument the full LST telescope camera.

\begin{figure}[t]
\centering
\includegraphics[width=3.4in]{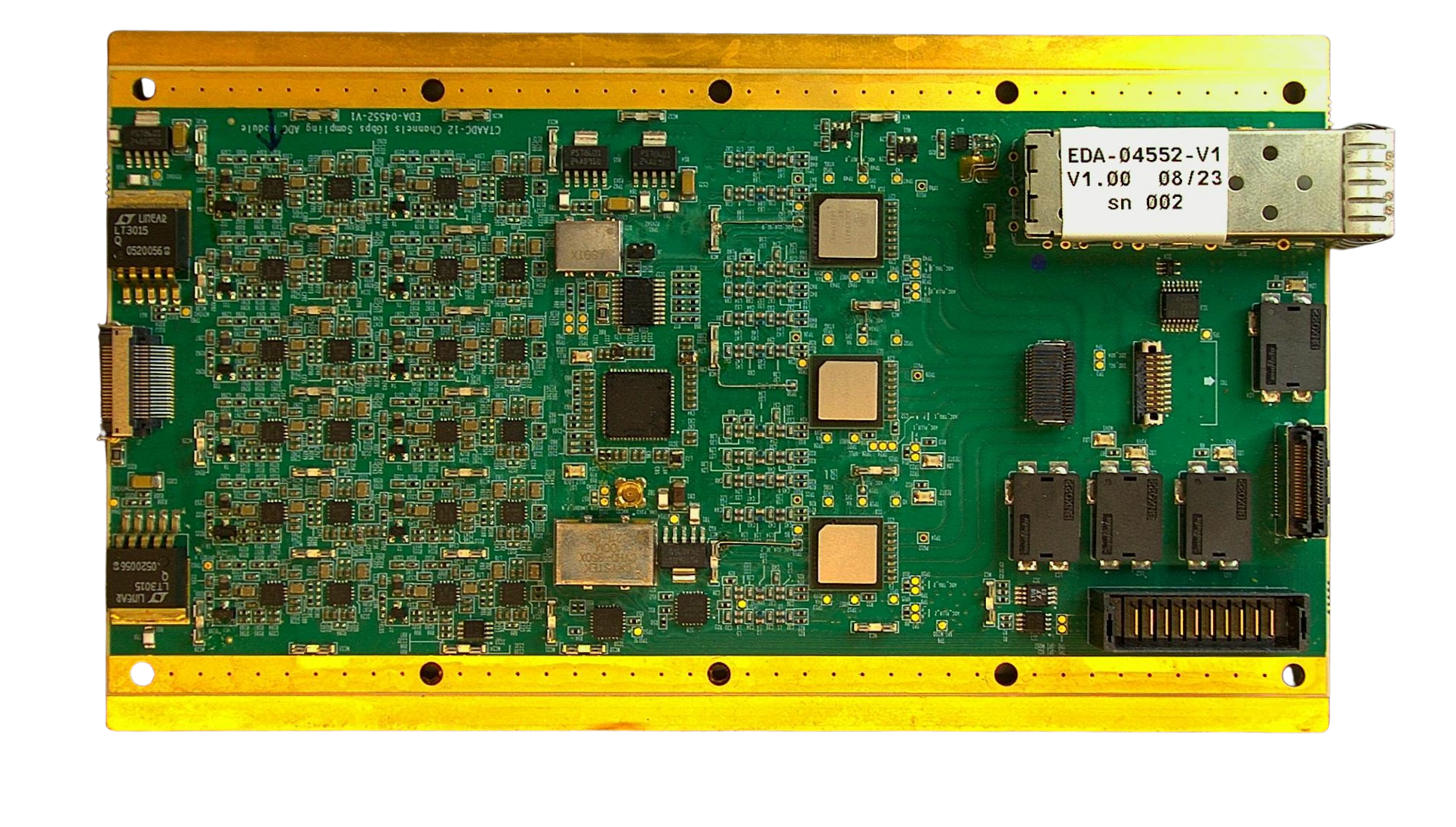}
\caption{The FADC board.}
\label{fadc}
\end{figure}

\begin{figure}[t]
\centering
\includegraphics[width=3.4in]{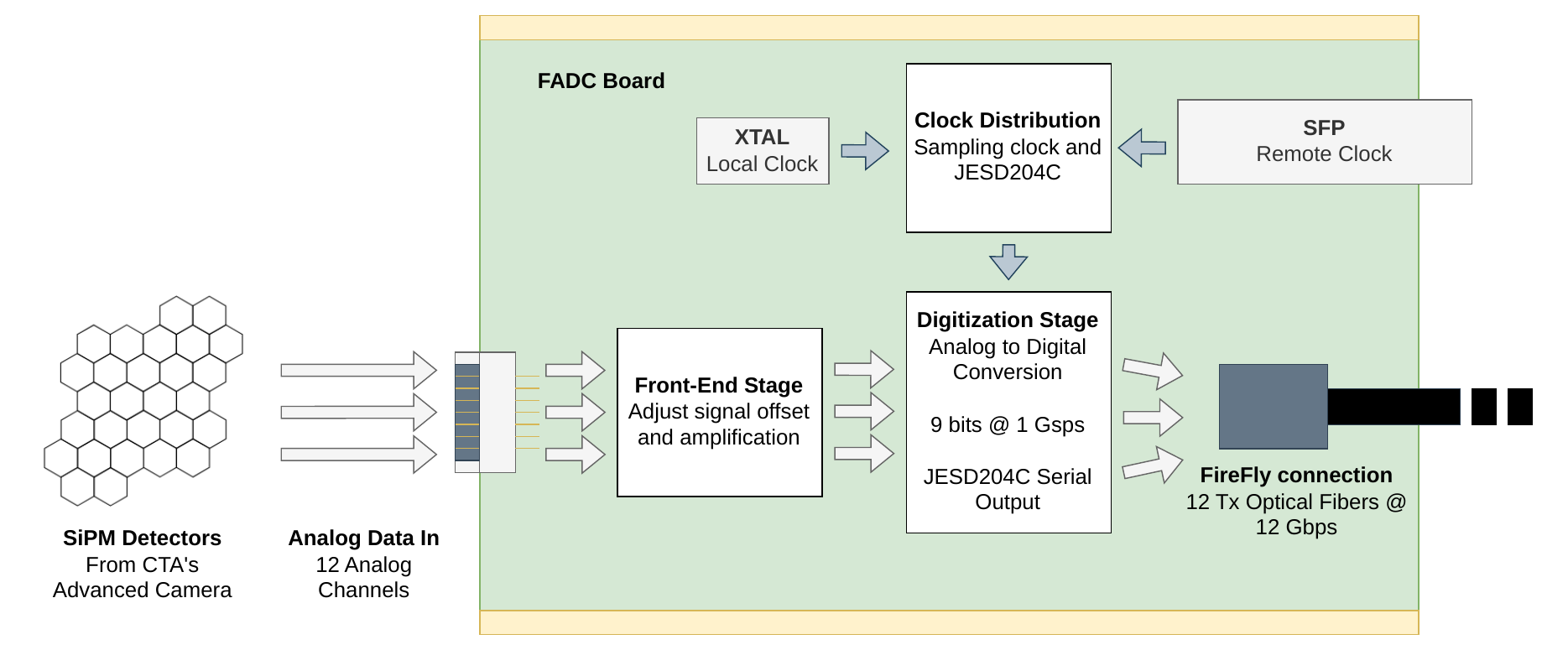}
\caption{Simplified block diagram for the FADC board.}
\label{fadc_block}
\end{figure}

This paper describes a proof-of-concept digital readout electronics system which manages the full acquisition chain, from the digitization of the sensors' waveform to their storage into the computer's farm memory. The digitization is performed by a custom designed board called \textit{FADC board}, designed to cope with typical SiPM signals. Its digital data are then sent via optical fibers to a second, FPGA-based, board whose job is to handle data reconstruction, event fragments storage, local trigger, and data acquisition. The novel approach is the FPGA implementation of Remote Direct Memory Access (RDMA) technology, which allows the use of commercial off-the-shelf Ethernet equipment to build the rest of the DAQ chain. Specifically, high-performance Ethernet switches are used to funnel several Back-End (BE) data links into a reduced number of higher-speed links that are connected to servers equipped with commercially available RDMA-capable Network Interface Cards (NIC). A representation of the overall system is visible in Fig.~\ref{sketch}.

Traditional Ethernet protocols, such as UDP, require the CPU to handle data packet processing and context switching, which often becomes a bottleneck for large data transfers typical in High-Energy and large-channel-count physics experiments \cite{juno-ipbus}. In contrast, RDMA protocols, although some may use UDP as transport layer (e.g., RoCEv2), offload packet processing and memory operations to dedicated hardware (RDMA-enable NICs), enabling zero-copy transfers and minimizing CPU involvement. This results in a low-latency/high-throughput data channel \cite{low-lat} ideal for high-performance computing, data analytics, and cloud computing environments.

\section{The FADC Board}
The FADC board, shown in Fig.~\ref{fadc} is a custom design electronics board measuring 165 cm x 95 cm, used to process analogue signals coming from 12 SiPM sensors and output the corresponding digitized data to BE electronics. The overall architecture is depicted in Fig.~\ref{fadc_block}, which illustrates the two main functional stages of the board: the analog pre-amplification stage and the digital signal processing stage. Each block in the figure corresponds to a key component or system in the board's architecture, as detailed below. For board configuration and analog signal interfacing, two Samtec connectors \cite{samtec_conn} are utilized. One connector is used to interface with an adapter board, allowing for Micro-miniature coaxial (MMCX) connection of the analog data from the SiPMs, while the other is used to link to an additional board that provides connections for configuration signals, including SPI and I\textsuperscript{2}C, and SubMiniature version A (SMA) connectors for the reference clock. 

\subsection{Analog Pre-Amplification Stage}
The analog data from the SiPMs undergoes an initial pre-amplification process. This stage has been designed with maximum flexibility in mind to accommodate the current lack of a dedicated pre-amplification Application Specific Integrated Circuit (ASIC) for the experiment's SiPMs. As the specific ASIC is in development by the LST-AdvCam collaboration, the board features a versatile pre-amplification system to facilitate testing and debugging. The stage includes:
\begin{itemize}
    \item \textbf{I\textsuperscript{2}C programmable Digital to Analog Converter (DAC)}: A DAC \cite{dac} is used to control the voltage offset of the input signal, providing precise control over the signal baseline. The DAC is user-manageable via an I\textsuperscript{2}C interface.
    \item \textbf{SPI Programmable Variable-Gain Amplifier (VGA)}: The amplifier \cite{vga} enables dynamic control of signal amplification, offering a wide range of gain from -6 dB to 26 dB. The SPI programmability ensures that the amplification can be tailored to the signal levels produced by the SiPMs, for an optimal interfacing with the dynamics of the ADCs.
    \item \textbf{Low-pass anti-alias filter}: To ensure that the signal is properly conditioned before digitization, a 5th-order low-pass anti-alias filter has been designed and included, with a typical cutoff frequency of 500 MHz, in accordance with the 1 Gsps (Giga-samples per second) sampling target rate of the ADCs.
\end{itemize}

\subsection{Digitization Stage}
\label{sec:digitization_stage}
The second stage is responsible for converting conditioned analog signals into digital data for further processing. The key components are the high-speed Analog to Digital Converters (ADCs). The FADC uses three quad-channel ADCs\cite{adc09qj1300}, each capable of sampling up to 1.3 Gsps with a resolution of 9 bits. To meet the experiment's requirements of performance and power, the ADCs are configured to run at 1 Gsps. However, these chips are footprint compatible with higher-resolution ADCs that can sample at 12 bits and 1.6 Gsps, offering future upgrade potential without requiring design changes to the board. To communicate with the downstream electronics, the converters implement the JESD204C protocol\cite{jesd204c}, a high-speed serial data transmission interface specifically designed for efficient, low-latency, and synchronized data transfer. The ADCs are driven by an ultra-low-noise Phase Locked Loop (PLL) system\cite{lmk04828}, compliant with JESD204C, which provides a 1 GHz clock for the digitization process. The PLL provides both clock generation and the SYSREF signals required for synchronization across the JESD204C links. The reference clock for the PLL can be provided via multiple sources:
\begin{itemize}
    \item Locally - on-board XTAL oscillator
    \item Remotely - Small Form-factor Pluggable (SFP) optical fiber input
    \item Remotely - SMA connector input
\end{itemize}
Once digitized, digital data are transmitted from the board via 12 JESD204C-compliant lanes. These lanes are routed to a 12 Tx Samtec FireFly connector \cite{samtec_ff}, which provides several benefits:
\begin{itemize}
    \item Mid-board placement: The placement of the FireFly connector mid-board improves signal integrity by minimizing signal path lengths, reducing potential noise attenuation and stray capacitance coupling.
    \item Compact optical fibers design: The system utilizes 12 optical fibers for data transmission, arranged in a ribbon fiber configuration. This arrangement contributes to the compactness of the design, enabling high data throughput while minimizing the space required for cabling.
    \item High bandwidth: The Samtec FireFly connector is capable of sustaining data rates well in excess of 12.375 Gb/s per lane, our target rate, making it well-suited for our application.
\end{itemize}

\subsection{Power Consumption}
\begin{figure}[t]
\centering
\includegraphics[width=2.4in]{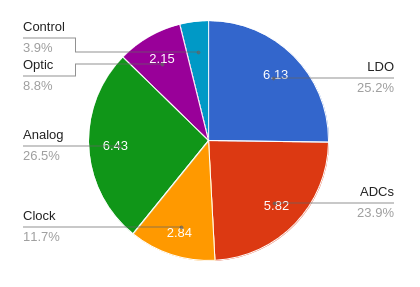}
\caption{Power consumption distribution for the FADC board.}
\label{pow_pie}
\end{figure}

As is typical for on-detector electronics, power consumption is severely constrained by area and thermal dissipation. Since the FADC is installed in the LST telescope camera, it is subject to the same stringent power requirements, which must therefore be carefully evaluated. In the design stage, the power simulations estimated a total power consumption of 24.3 W, with the different contributions outlined in Fig.~\ref{pow_pie}. The actual power consumption of the board is about 21 W. The difference is likely due to the slower sampling frequencies used by the ADCs, 1000 Msps, as the datasheets report the power consumption for the maximum digitization rate, 1300 MSample Per Second (Msps).
It is worth noting that the analog pre-amplification stage represents about 26\% of the total power consumption. Once the current amplification stage is replaced in favor of the dedicated pre-amplifier ASIC, the analog power consumption is expected to decrease substantially.

Care has been taken in the design of the PCB to facilitate thermal dissipation: copper thickness of central ground planes has been increased to 70 $\upmu$m and, thanks to a special manufacturer technique, those planes protrude to the outside of the board perimeter so that very efficient thermal anchoring to the surrounding structures is made possible.

\subsection{Failure In Time (FIT) Analysis}
Reliability assessment is a fundamental aspect of electronic system design, particularly in safety-critical and long-lifetime applications. Two commonly used metrics to quantify the reliability of electronic components are Failure In Time (FIT) and Mean Time Between Failures (MTBF). The FIT rate, expressed in failures per billion hours ($10^9$ device-hours), provides a standardized measure of the expected failure rate under specified operating conditions. Conversely, MTBF, typically given in hours, represents the average operational time between successive failures and is inversely proportional to the FIT rate. These metrics are derived from statistical failure models, often assuming an exponential distribution of failure events during the useful life period of a component.   

For the AdvCam, such reliability analysis is particularly critical due to the need for long-term operation with minimal maintenance, where any unexpected board failure may require halting observations and performing difficult on-site interventions. FIT-based evaluations help inform component selection, thermal design, and redundancy planning to ensure continuous operation throughout the expected lifetime of the observatory.

Based on an open-source work \cite{jochen}, a custom \textit{C++} software has been developed to estimate the overall FIT of the FADC board based on the components used and their specific parameters. A critical parameter in FIT calculations is temperature, as it significantly impacts component reliability. While the operating temperature of the board is relatively uniform, certain components, such as power regulators and heavy loads (like ADCs) experience elevated temperatures due to increased power dissipation. To accurately estimate the temperature differential between these critical components and the ambient board temperature, thermal camera images were acquired under operational conditions. 

All component data was collected with a 60\% confidence level, a standard industry practice. FIT data is typically provided at a standard temperature of 55°C. To account for these temperature differences, a temperature derating formula was applied to adjust the FIT rates. Other factors that would normally need derating (power dissipation, electrical stress, mechanical stress, etc.) have been ignored as their contribution is small in comparison and can be considered negligible for our purposes, as long as the component is used within specifications. To apply temperature derating the following formula is used:
\begin{equation}
    \lambda_{part} = \lambda_{ref} \times \pi_t
\end{equation}
 where $\pi_t$ is the thermal derating coefficient derived from \textit{Arrhenius' law}:
\begin{equation}
    \pi_t = exp \left[ \frac{E_a}{k_b}\left( \frac{1}{T_{ref}} - \frac{1}{T_{op}} \right) \right]
\end{equation}
where $E_a$ is the defect activation energy, $k_b$ is the Boltzmann's constant, $T_{ref}$ is the reference temperature (typically 55°C) and $T_{op}$ is the operating temperature.

\begin{figure}[t]
\centering
\includegraphics[width=2.8in]{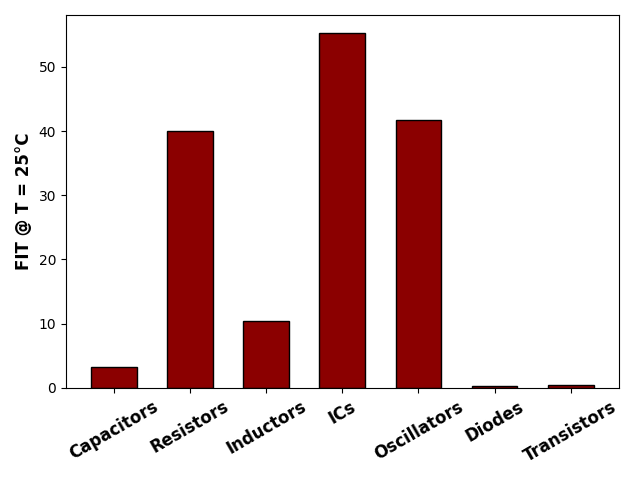}
\caption{Contributions to the total FIT at 25°C.}
\label{fit_contributors}
\end{figure}

The estimated total FIT at 25°C is about 150, with the different contributions shown in Fig.~\ref{fit_contributors}. The atypical big contribution to the FIT given by the resistors is mostly due to a current sense resistor, the TL3AR005F, for which the reference FIT is 159 at 70°C. Having four of these mounted on the FADC, their total contribution after the temperature derating is 38.4 at 25°C, which accounts for about 96\% of the total resistor contribution.
Another significant contribution is given by the oscillators. Two oscillators, with reference FITs of 250 @ 55°C and 116 @ 45°C, collectively account for approximately 30\% of the total FADC FIT. Identifying the major contributors to the overall FIT, such as the oscillators and current sense resistors, is important for a future version of the FADC, where targeted design improvements can be made to enhance the system's reliability. Elements excluded from FIT computation are connectors, for which the reliability is expressed in numbers of guaranteed mating cycles, the bare PCB, where the failure rate, which depends on the complexity (number of layers, type of vias, etc.) and fabrication details, is usually not provided by the manufacturers, and PCB assembly.

\begin{figure}[t]
\centering
\includegraphics[width=3.2in]{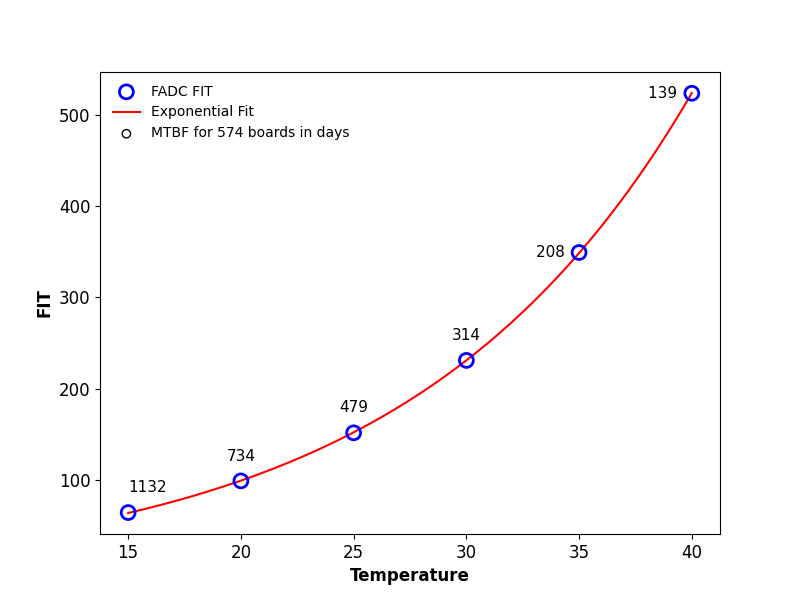}
\caption{Dependency of the FADC FIT with temperature. MTBF for 574 boards is also expressed for each point in days.}
\label{fit_temp}
\end{figure}

Fig.~\ref{fit_temp} shows the dependency of the overall FIT with the temperature. The plot also shows the MTBF in days for the 574 foreseen number of boards to be used in the AdvCam; notably, it indicates that, if maintained at 20\degree C, a single board failure—and thus replacement—is expected approximately every two years. 

\subsection{Simulation and Tests}
To ensure the performance and reliability of the board, extensive tests were carried out, particularly focusing on the characterization of the high-speed digital data lanes and the clock. Signal integrity assessments were performed using advanced simulation tools and direct measurements to verify the board's ability to handle high-speed data transmission effectively.

\subsubsection{Signal Integrity Simulations}

\begin{figure}[t]
\centering
\includegraphics[width=2.8in]{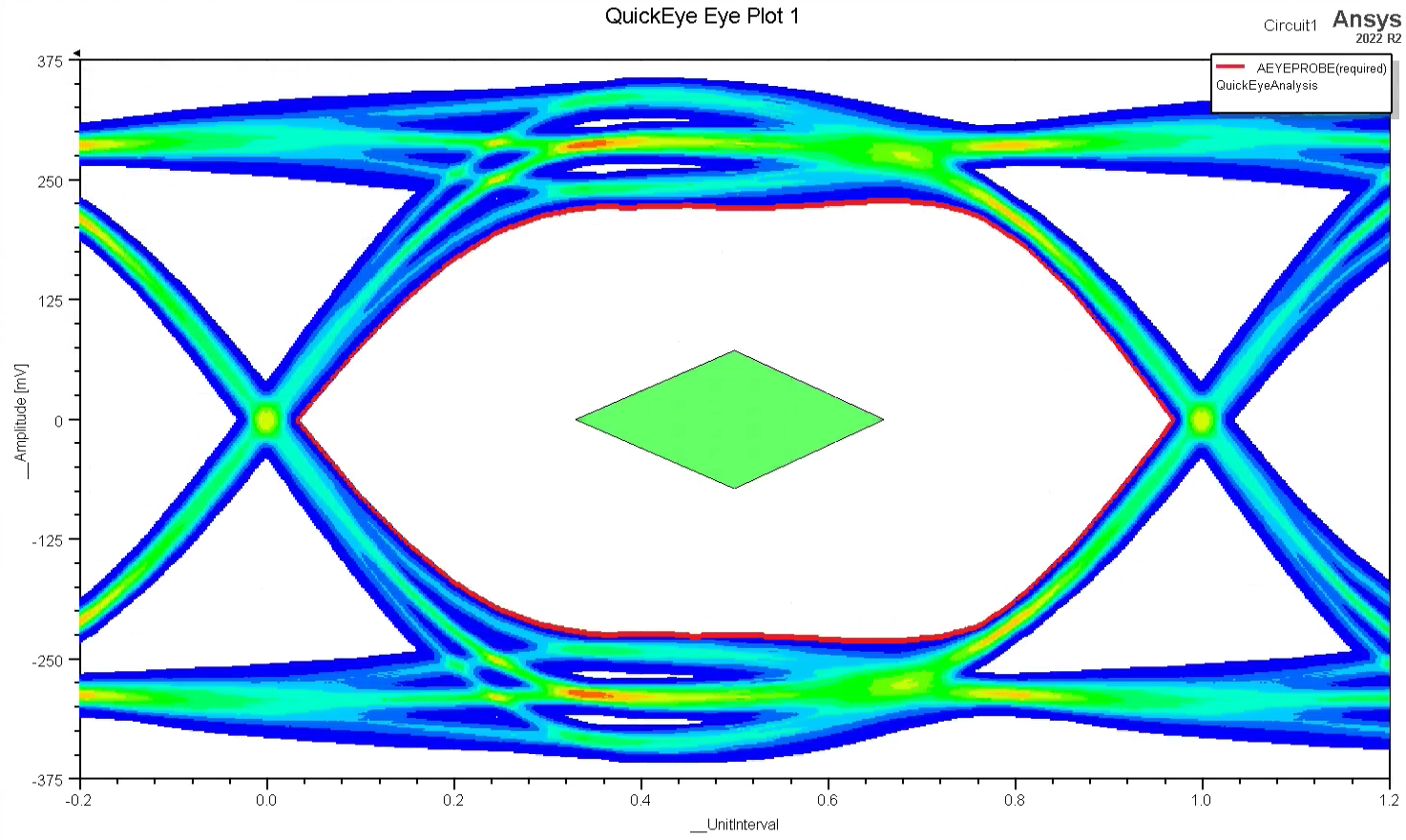}
\caption{Eye diagram simulation for the FADC's 12 Gb/s links generated with Ansys SIwave. The $10^{-12}$ BER mask is over-imposed in green.}
\label{sim_eye}
\end{figure}

The high-speed lanes, which operate at about 12.5 Gb/s per lane, were thoroughly evaluated using the Ansys SIwave software application\cite{ansys}, together with Cadence SPB Allegro \cite{cad_ence} for the Printed Circuit Board (PCB) CAD model. The tool is a 3D field solver for the design and analysis of high-speed PCBs and integrated circuit (IC) packages. Its capabilities are particularly suited for addressing complex electromagnetic, signal integrity and power integrity issues that arise in high-frequency circuits. For the fast lanes, SIwave computed the S-parameters of the tracks  accounting for the shape, dielectric material, vias geometry and board stackup, yielding a model to be used with a Spice simulator for Bit Error Rate (BER) and jitter estimation.
As shown in Fig.~\ref{sim_eye}, the eye diagram generated for the high-speed links shows a clean, wide open eye with minimal noise and jitter. The green area inside the eye represents the mask for the worst case optical eye tolerated by the optical receiver, in order to achieve a BER less or equal than $10^{-12}$, which is standard for data transmission over fiber optics and Ethernet channels. As showed, the simulated eye significantly exceeds the requirements imposed by the mask.

\subsubsection{High-Speed Probing and Measurements}
\begin{figure}[t]
\centering
\includegraphics[width=3.4in]{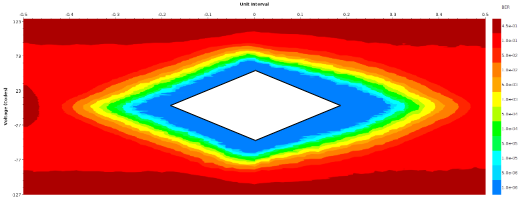}
\caption{Eye diagram for the FADC's 12 Gb/s links retrieved using Xilinx/AMD In-system IBERT tool. The $10^{-12}$ BER mask is over-imposed in white.}
\label{eye}
\end{figure}

Following the simulations, physical measurements were carried out to further validate the performance of the high-speed lanes. By employing a Xilinx/AMD KCU105 evaluation board  with a FireFly-to-FMC adapter, the in-system IBERT Xilinx/AMD IP core \cite{ibert} was used to measure and generate the eye diagram directly from the JESD204C 12 Gb/s links, shown in Fig.~\ref{eye}. Differently from the simulation, this eye diagram not only considers the FADC's PCB trace from the ADCs to the FireFly, but it also incorporates the effects of the FireFly modules themselves, the optical fibers, the FMC adapter, the KCU105 PCB and, finally, the FPGA transceiver. Nevertheless, these measured values confirm that the board's high-speed communication lanes are operating well within performance specifications.

\subsubsection{ADC sampling clock quality assessment}


The performance of high-speed ADCs is strongly influenced by the quality of the sampling clock, particularly its phase noise and jitter characteristics. To evaluate the quality of the clock used in our 1 GHz sampling system, we measured the clock period stability using a spectrum analyzer set up as a phase noise analyzer. The analyzer integrated the phase noise to determine the jitter value. Following the PLL datasheet, the frequency range for analysis was set from 100 Hz to 30 MHz with spurs removed. In these conditions, the random jitter was estimated at around 770 fs RMS.

While excessive jitter degrades the ADC's Signal-to-Noise Ratio (SNR) by introducing additional timing uncertainty, its impact depends on the input signal frequency. The contribution of the sampling clock jitter to the total SNR, defined as Signal-to-aperture-Jitter-Noise-Ratio (SJNR) \cite{adc-testing}, is given by the following formula \cite{snr_jitter}:
\begin{equation}
    \text{SJNR}_{[dB]} = 104_{[dB]} - 20 \cdot \log\left(\frac{f_{in}}{1_{[MHz]}}\right) - 20 \cdot \log\left(\frac{\sigma_{\text{jitter}}}{1_{[ps]}}\right)
    \label{eq:snr_jitter}
\end{equation}
where $f_{in}$ is the input analog frequency and $\sigma_\text{jitter}$ is the RMS jitter of the sampling clock.

The ADC's datasheet specifies a SNR of 53.5 dBFS for an input frequency of 100 MHz \cite{adc09qj1300}. Using Eq.~\ref{eq:snr_jitter}, the calculated SJNR is 66.3 dB. This indicates that the performance of the ADC is primarily limited by other noise sources, such as thermal or quantization noise, rather than the sampling clock jitter. 

Although further improvements are possible, as a comparative analysis with the PLL evaluation board reveals that our clock exhibits approximately $3\times$ higher jitter, the $\backsim 770$ fs jitter from the PLL does not constitute the dominant performance-limiting factor in this system.

\subsubsection{SINAD and ENOB Characterization}
\label{sec:enob}
To evaluate the performance of the digitizer front-end board, the Signal-to-Noise and Distortion Ratio (SINAD) and Effective Number of Bits (ENOB) of the digitized data were measured. For the measurement, a high-purity sine wave signal, assumed to be ideal, at 200 MHz was generated using a Keysight M8190A 12 GSa/s arbitrary waveform generator \cite{wave_gen}. The signal amplitude and the analog front-end of the board was set to span the full dynamic range of the ADC. The digitized output data were acquired by connecting the FADC board to the Xilinx/AMD KCU105-based BE electronics system described in the following section, which interfaces with a PC through a 10 Gb/s Ethernet connection.

The acquired data were analyzed using the Python library pysnr~\cite{pysnr}, which provides functionality analogous to MATLAB's \textit{sinad} command. This function determines the SINAD using a modified periodogram of the same length as the input signal. 

\begin{figure}[t]
\centering
\includegraphics[width=3.4in]{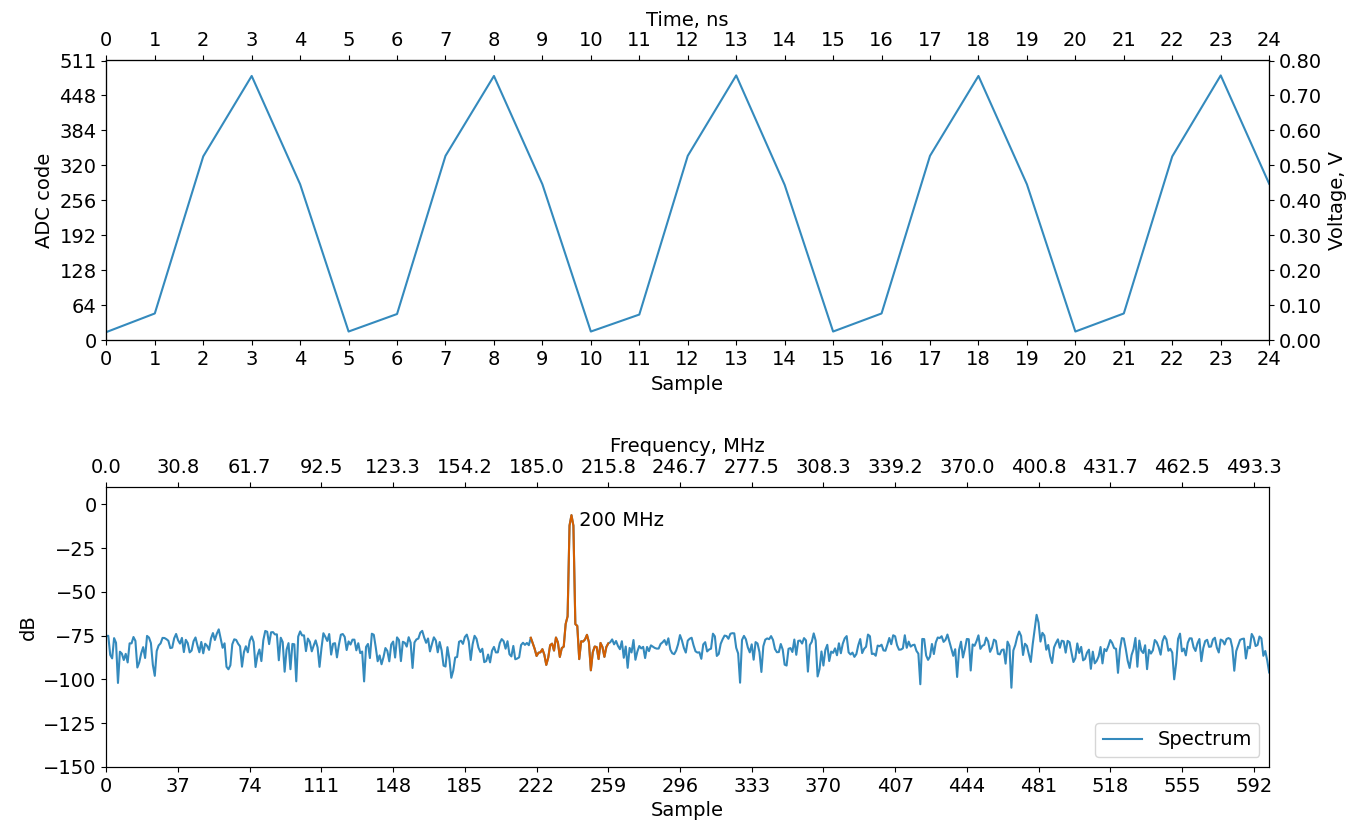}
\caption{Data and FFT spectrum for SINAD/ENOB calculation.}
\label{sinad-adceval}
\end{figure}

The acquired data with the relative Fast Fourier Transform (FFT) spectrum are visible in Fig.~\ref{sinad-adceval}. As shown, the measured SINAD was 48.8 dB, corresponding to an ENOB of approximately 7.8 bits.

\subsubsection{Crosstalk}
To further evaluate the performance of the digitizer front-end board, the crosstalk between adjacent channels was measured. Crosstalk, defined as the unintended coupling of a signal from one channel (aggressor) to another (victim), was characterized in terms of isolation in dB.

For this measurement, the same setup used in Sec.~\ref{sec:enob} was used: high-purity sine waves signal at different frequencies were generated using a Keysight M8190A 12 GSa/s arbitrary waveform generator \cite{wave_gen}. The signal amplitude and the analog front-end of the board were set to span the full dynamic range of the ADC. The signal was applied to one channel (aggressor), while an adjacent channel (victim) was left unconnected, terminated with a matched impedance to minimize reflections.

The acquired data from both the aggressor and victim channels were analyzed using the Fast Fourier Transform (FFT) to evaluate the spectral content of the victim channel and quantify the level of induced interference. The measured crosstalk was determined by comparing the amplitude of the signal present in the aggressor channel with the unwanted signal amplitude detected in the victim channel. The crosstalk isolation, calculated as:

\begin{equation}
XT = 20 \log_{10} \left( \frac{V_{\text{victim}}}{V_{\text{aggressor}}} \right)
\end{equation}

where $V_{\text{victim}}$ and $V_{\text{aggressor}}$ are the signal amplitudes in the victim and aggressor channels, respectively, resulted in an average value of approximately 45 dB across the tested frequency range.

\subsubsection{Linearity}
\begin{figure}[t]
\centering
\begin{subfigure}{.24\textwidth}
  \centering
  \includegraphics[width=1.0\linewidth]{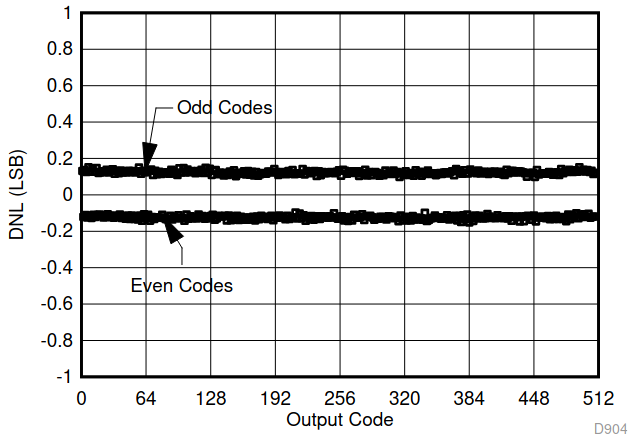}
  \caption{}
  \label{dnl_ds}
\end{subfigure}%
\begin{subfigure}{.24\textwidth}
  \centering
  \includegraphics[width=1.0\linewidth]{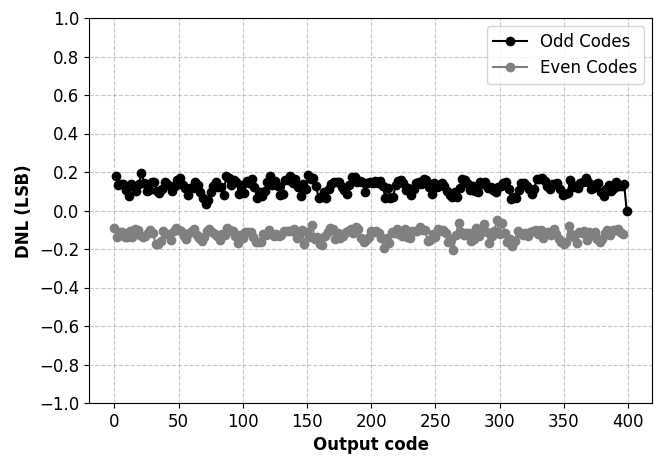}
  \caption{}
  \label{dnl_lab}
\end{subfigure}
\caption{DNL vs Code in the ADC datasheet (a) and FADC (b).}
\label{dnl}
\end{figure}

\begin{figure}[t]
\centering
\begin{subfigure}{.24\textwidth}
  \centering
  \includegraphics[width=1.0\linewidth]{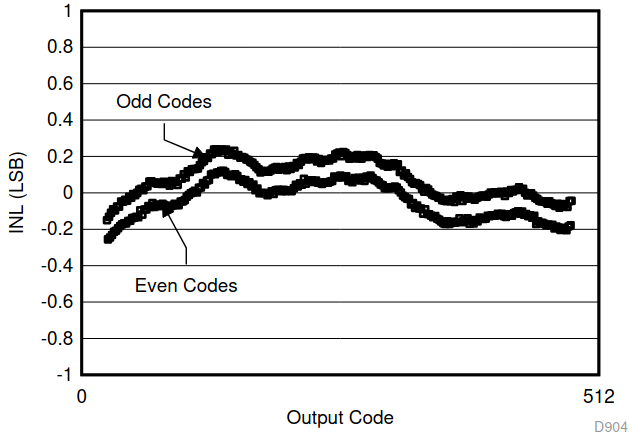}
  \caption{}
  \label{inl_ds}
\end{subfigure}%
\begin{subfigure}{.24\textwidth}
  \centering
  \includegraphics[width=1.0\linewidth]{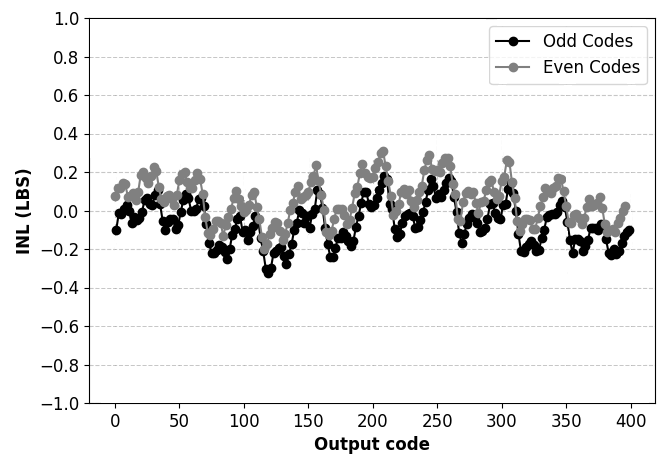}
  \caption{}
  \label{inl_lab}
\end{subfigure}
\caption{INL vs Code in the ADC datasheet (a) and FADC board (b).}
\label{inl}
\end{figure}

To assess the linearity of the analog-to-digital conversion in the FADC system, we followed the methodology outlined in \cite{linear}, which employs a sinusoidal input signal to derive the Differential Non-Linearity (DNL) and Integral Non-Linearity (INL) metrics. The sinusoidal input, provided by a SRS-DS360~\cite{srs} is assumed to be ideal.

A comparison between the measurements from our setup with the ADC datasheet, shown in Fig.~\ref{dnl} for the DNL and Fig.~\ref{inl} for the INL, indicates a close match between the two, with our measurements exhibiting a slightly larger variance, especially for the INL, causing the data points to be more dispersed around the mean. 

This increased variance is likely attributable to the influence of the analog front-end used for signal conditioning prior to digitization. Despite this difference, the overall magnitude of the linearity deviations remains comparable, suggesting that the signal conditioning stage introduces only minor additional non-linearity effects.

\section{Back-End Electronics System}
A BE system mock-up has been developed to serve as a functional prototype, allowing us to test and validate our FE electronics and data acquisition processes. It is based on the Xilinx/AMD KCU105 development platform \cite{kcu105}, with the FPGA tasked with receiving, processing and storing the data. In addition, the firmware implements a trigger mechanism to control data selection.

The board interfaces with the FADC through high-speed optical fibers. One pair is reserved for the reference clock using the SFP option on the FADC board, as described in Sec.~\ref{sec:digitization_stage}, while the remaining 12 fibers receive the JESD204C-formatted data at about 12 Gb/s.  To address the critical need for synchronization in multi-board systems, the design can accommodate a White Rabbit\cite{white} FMC Mezzanine\cite{cute-wr, shine}. This mezzanine card provides highly accurate timing distribution, ensuring coherence across all boards. Alternatively, other synchronization schemes based on IEEE 1588, which specifies the Precision Time Protocol (PTP), \cite{ieee1588} can also be evaluated for systems where White Rabbit may not be applicable or required \cite{nanosec}.

One of the key aspects of the FPGA firmware is the use of RDMA technology to efficiently transfer the triggered data off the board and onto connected servers for further analysis. Specifically, the board leverages RoCEv2 implemented in the FPGA hardware, at a transmission rate of 10 Gb/s. Further sections of this paper will detail the firmware design, along with performance results and future developments.

\subsection{DAQ Firmware}
\begin{figure}[t]
\centering
\includegraphics[width=3in]{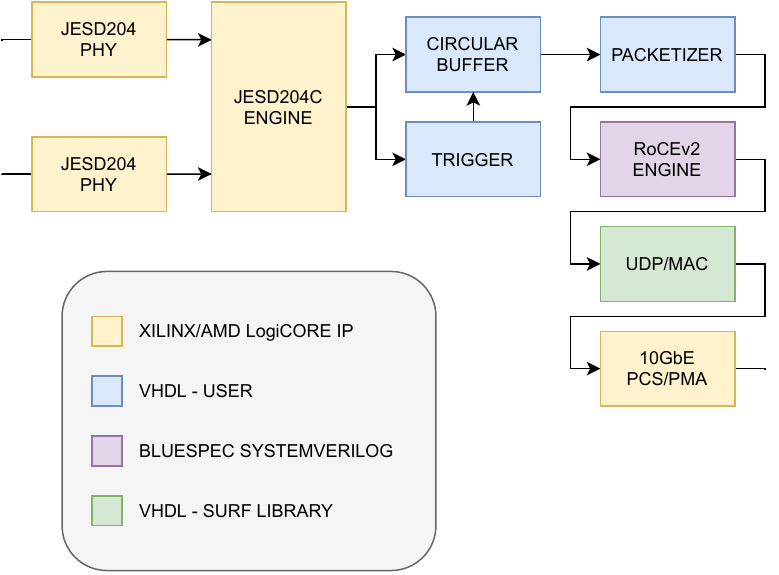}
\caption{BE firmware block diagram.}
\label{firmware-block}
\end{figure}

As mentioned above, the main task of the BE FPGA is to retrieve the JESD204C event fragments data from the FE, store them in memory, run a local trigger algorithm, and ship the accepted data to a server. Fig.~\ref{firmware-block} illustrates the block diagram outlining the various steps involved in the firmware. The communication bus represented by the arrows is AXI4-Stream, with data widths between 64 and 256 bits. 

The JESD204C formatted data by the sampling ADCs get recovered in the FPGA using the related Xilinx/AMD LogiCORE IP. The IP receives the high-speed lanes as input and provides as output the recovered data in groups of 64 bits clocked at 187.5 MHz.
The recovered data are then duplicated: one copy gets stored in a circular buffer while the other copy is analyzed for a trigger decision. The trigger algorithm implemented in the FPGA is a basic leading-edge trigger, while in production firmware it will be upgraded by another trigger algorithm, being developed by the LST-AdvCam collaboration, that relies on the digital sum of the cluster pixel signals and its direct neighbors\cite{l1board}. This reduced version is solely required for testing, and modifications will impact only on the data rate at the output of each BE board.

As soon as the digital signal is found to be above a programmable threshold, a 50 ns window of data related to the trigger timestamp is drawn from the circular buffer and sent over to the packetizer module. This module is responsible for interfacing with the RoCEv2 engine.

\subsection{RoCEv2 - RDMA Core}
The RoCEv2 \cite{rocev2} core has been written in Bluespec SystemVerilog (BSV)\cite{bsv} to operate as a full featured Host Channel Adapter (HCA). It enables RDMA over Ethernet for low-latency, high-throughput data transfers without CPU intervention or requirement. The core implements RDMA functionalities, such as Queue Pair (QP) management, reliable transport, and RDMA operations (e.g., SEND, RDMA WRITE, RDMA READ) over standard Ethernet/IP networks. It integrates with the FPGA fabric to handle packet processing, memory registration, and DMA operations. The core only needs interfacing with the system’s memory, Ethernet transport layer, PHYs and RoCEv2 Connection Manager. This configuration has many beneficial consequences: it allows the FPGA to serve as a programmable RDMA endpoint in different roles, e.g. for high-performance computing, storage, networking applications or custom electronics; it enables speed scaling from 1Gb/s to 100 Gb/s or beyond (e.g. 400 Gb/s) depending on the PHY type and FPGA technology. Testing the core performances, we targeted a Xilinx/AMD Virtex Ultrascale Plus VU9P device with the number of QPs resticted to one and exchanged RDMA packets at 100Gb/s with a Mellanox ConnectX-5 in a commodity server; it facilitates the embedding of RDMA functionality - at least partially - to severely constrained environments where large memories for tens or hundreds of QPs may not be available or not suitable, as is the case, for example, of readout electronics in high radiation environments. It is worth noting also that the open-source BSV compiler, bsc\cite{bsc}, generates plain Verilog code with no dependencies: this is ideal for targeting different ASIC technologies, depending on the Ethernet line rate chosen and on the availability of a suitable MAC.

The RoCEv2 standard details a protocol\cite{rocev2} derived from InfiniBand\cite{infiniband} designed for efficient data transfer in data centers. It defines roles for Host Channel Adapters (HCAs), typically implemented as NICs in servers, and Target Channel Adapters (TCAs), which may have reduced capabilities, such as in network-attached storage or, in our case, custom readout electronics in an astrophysics experiment. By noting that —  as often is the case in physics experiments — the natural flow of data is from the detector to the processing software, it is straightforward to conclude that our implementation of a TCA may support the RDMA operation only in one direction, e.g. RDMA WRITE, and operate without loss of functionality and interoperability with commercial off-the-shelf HCAs.

\subsection{RoCEv2 - RDMA Firmware}
The FPGA implementation of the RoCEv2 core is based on an open-source project\cite{blue-rdma}, developed using BSV by two of the authors. BSV is a high-level hardware description language that facilitates the design of complex hardware systems by abstracting low-level details, allowing for more modular and scalable designs. Compared to High-Level Synthesis (HLS), which typically converts C/C++ code into hardware, BSV offers more precise control over hardware structures while maintaining a higher level of abstraction than traditional Hardware Description Languages (HDL)\cite{hlsVsBsv}. 
An additional advantage of using a BSV-based design over similar projects implemented with Vivado HLS\cite{ultrasound, atlas} is the portability of the code, as Vivado HLS restricts FPGA targets to only Xilinx/AMD devices. This flexibility allows BSV-based designs to be used in specialized environments, such as high-energy physics experiments or space applications, where radiation-tolerant or radiation-hardened FPGAs are required for high radiation environment\cite{obdt, radiation}.

The original core has been modified by removing the entire data reception path as well as support for all RDMA READ operations. The rationale behind this operation is not only the nature of the application but also due to cost reasons: the AdvCam is made of about 7000 channels to sample, readout, process, store and broadcast, and FPGA area will seriously impact the overall cost per channel. As shown in Tab.~\ref{roce-comparison}, the modified core results in a lighter and more portable design suitable for deployment on smaller FPGAs.
Moreover, the RoCEv2 core guarantees reliability in data transfer. By using it in Reliable Connection (RC) mode, the core provides features such as acknowledgment and retransmission, ensuring that data are delivered reliably and in order, making up for the intrinsic unreliability of the UDP layer. 
RC mode is mandatory in physics experiments applications because data fragments from different parts of the detector are funneled at random times through network devices (e.g. RoCEv2 endpoints and switches) to DAQ servers for event building: this process cannot rely on a bare UDP transport which is lossy by construction and is unable to handle potential congestion of switch ports \cite{congestion}.

\begin{table}[!t]
\renewcommand{\arraystretch}{1.3}
\caption{Comparison of RoCEv2 resource estimates after synthesis between original and modified core. The target is a Xilinx/AMD XCKU040 FPGA}
\label{roce-comparison}
\centering
\begin{tabular}{l|l|l|}
\cline{2-3}
                                                  & \cellcolor[HTML]{EFEFEF}Original & \cellcolor[HTML]{EFEFEF}Modified \\ \hline
\multicolumn{1}{|l|}{\cellcolor[HTML]{EFEFEF}Look Up Tables (LUT)} & 92475 {[}38.2\%{]}               & 29802 {[}12.3\%{]}               \\ \hline
\multicolumn{1}{|l|}{\cellcolor[HTML]{EFEFEF}Flip Flops (FF)}  & 136599 {[}28.2\%{]}               & 40902 {[}8.4\%{]}               \\ \hline
\multicolumn{1}{|l|}{\cellcolor[HTML]{EFEFEF}Block RAM Tile}  & 18 {[}3.0\%{]}               & 8.5 {[}1.4\%{]}               \\ \hline
\end{tabular}
\end{table}

A candidate system design for AdvCam readout addresses channels in groups of 48, all processed by a single FPGA. The estimated transfer rate after zero suppression, local trigger processing and global trigger validation is largely compatible with 10 Gb/s Ethernet line rate. For this reason the R\&D on RDMA based readout has focused on this type of network.

As RoCEv2 uses UDP as its transport protocol, a full UDP/MAC network stack has been implemented. The 10 GbE network system is heavily based on the SLAC Ultimate RTL Framework (SURF)\cite{surf}, which consists of an open-source modular framework designed to facilitate data acquisition in FPGA-based systems\cite{surf-exp}. The framework integrates VHDL-based hardware modules and IPs with the SLAC ROGUE software\cite{rogue}, which provides a high-level software interface for data control, monitoring, and transfer over Ethernet. This integration allows for easy communication between the FPGA hardware and the software, enabling flexible and scalable data acquisition systems. Importantly, SURF also provides an additional layer called RUDP (Reliable UDP), which, together with its ROGUE software counterpart, ensures reliable data transfer. However, when transferring RoCEv2 traffic, this RUDP layer is bypassed, as the RoCEv2 core itself handles reliability.

In order to fully support the RoCEv2 protocol, the firmware is running a modified version of the SURF's UDP and MAC modules, with the main adjustments focused on enabling the insertion and validation of the iCRC field of the RoCEv2 packet. The modifications integrate seamlessly with the previous version, remaining transparent to the end-user when RoCEv2 packets are not flowing. These changes are now part of the official distribution of SURF. The connection management of the RoCEv2 core, including tasks such as creating a Protection Domain (PD), setting up Queue Pairs (QP), and managing protection keys, is fully implemented in software exploiting the reliable register-access capability of the SURF/ROGUE network framework. For time-critical operations, such as memory address management and target IP reconfiguration, a dedicated Finite State Machine (FSM) is implemented in hardware to ensure low-latency control and response.

\begin{figure}[t]
\centering
\includegraphics[width=2.8in]{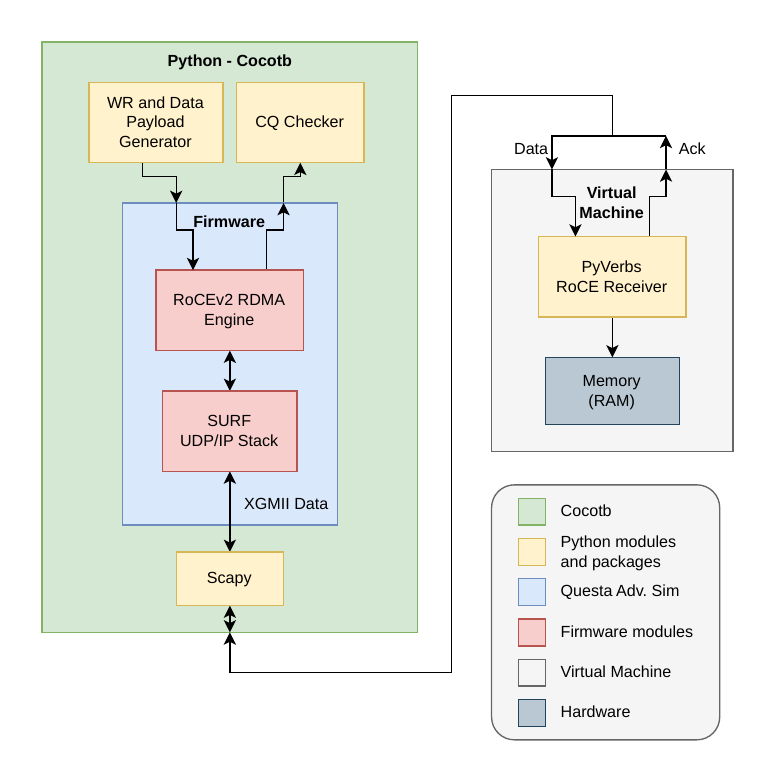}
\caption{RoCEv2 firmware simulation block diagram.}
\label{firmware_sim}
\end{figure}

\subsection{RoCEv2 Firmware Simulation}
The simulation environment for the FPGA implementation of the RoCEv2 protocol uses the UDP / MAC network stack provided by the SLAC SURF framework. As Fig.~\ref{firmware_sim} shows, the testbench for the firmware simulation, running with the Siemens Questa Advanced simulator engine \cite{questa}, is embedded within a Python environment using Cocotb, an advanced Python-based coroutine-driven simulation library to verify HDL designs \cite{cocotb}.

As foreseen in the final implementation, the simulation setup focuses on exercising the RDMA WRITE operation over RoCEv2. Within Cocotb, the input stimuli for the RoCEv2 engine are generated, by configuring the necessary elements of the RDMA system, such as the PD and QPs, as well as generating and issuing Work Requests (WR) with related payload data. The SURF UDP and MAC modules generate XGMII-formatted data, which is intercepted and processed using the cocotbext-eth Python extension to construct complete Ethernet frames. Using the Scapy Python library, these frames are then transmitted to a target virtual machine that runs a software-based RoCEv2 implementation (Soft-RoCE). On the virtual machine, a Python script that uses the Pyverbs library (a Python API over rdma-core, the Linux userspace C API for the RDMA stack) is responsible for handling the incoming RoCEv2 traffic.
As the connection mode between the simulated FPGA firmware and the virtual machine is RC, the virtual machine acknowledges the received RDMA data by sending response packets back to the firmware simulation. These acknowledgments are intercepted by Cocotb again using Scapy, where they are reinjected into the firmware simulation to update the RoCEv2 Completion Queue (CQ). The status of the CQ is then verified in Python to ensure that all WRs have been successfully completed.

Additionally, to validate data integrity, the memory region in the virtual machine is examined post-simulation to confirm that the data transmitted via RDMA WRITE have been correctly written to the target buffer. Using this method it was shown that, through a comprehensive verification flow, the RoCEv2 firmware operates as expected.

\subsection{Performance Tests}
After successfully implementing the RoCEv2 firmware on a Xilinx/AMD KCU105 evaluation board, a series of performance tests were conducted to evaluate its performance in different configurations, initially using a Soft-RoCE receiver and later a hardware-based set-up. To accurately measure the throughput in both scenarios, a custom software was developed. The software measures the time elapsed from the moment the first byte of payload data was written to the receiver's memory until the last byte was successfully transferred. Additionally, it verifies that all memory addresses between the first and last have been written to successfully, ensuring accurate measurement.
The total amount of data transferred in each test was approximately 8 GB, to ensure that the measured time difference was large enough to minimize the impact of any inherent timing inaccuracies in the software.
The throughput (T) was then calculated as:
\begin{equation}
    T = \frac{\text{Total Data Transferred (Bytes)}}{\text{Time Taken (seconds)}}
\end{equation}

\subsubsection{Test 1: PC with a Soft-RoCE Receiver}
\begin{figure}[t]
\centering
\includegraphics[width=3.3in]{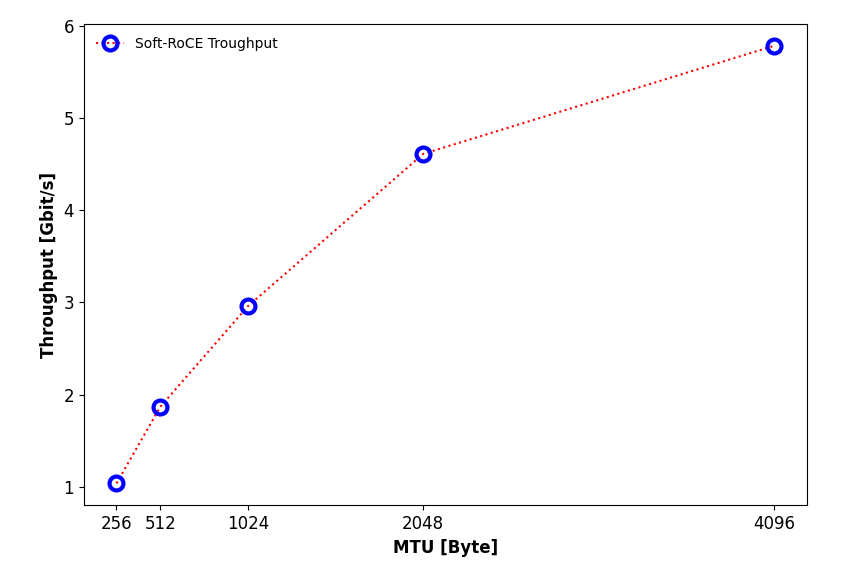}
\caption{Throughput as a function of MTU size for RoCEv2 communication between the KCU105 and a PC with a Soft-RoCE receiver.}
\label{soft-roce-throughput}
\end{figure}

In the first test, the KCU105 was connected to a standard PC running the same Soft-RoCE receiver used in the firmware simulation. 
Fig.~\ref{soft-roce-throughput} illustrates the relationship between the Maximum Transmission Unit (MTU) size and the achieved throughput. As expected, larger packet sizes result in higher throughput, as larger packets mitigate the effect of the overhead associated with packet headers. Given that latency was not a primary concern, the maximum possible MTU size of 4096 bytes was configured for RoCEv2 traffic.
In this configuration, the RoCEv2 implementation achieved a maximum throughput of approximately 5 Gb/s over the 10 Gb/s Ethernet link. This performance, while functional, was below the expected throughput for the link's capacity. It is likely that the bottleneck in this scenario was the receiving PC.
To confirm it, a second test was conducted using a server equipped with a Mellanox/NVIDIA ConnectX-5 NIC, which supports RDMA hardware offloads, eliminating the CPU bottleneck.

\subsubsection{Test 2: Server with Mellanox ConnectX-5 NIC}
\begin{figure}[t]
\centering
\includegraphics[width=3.3in]{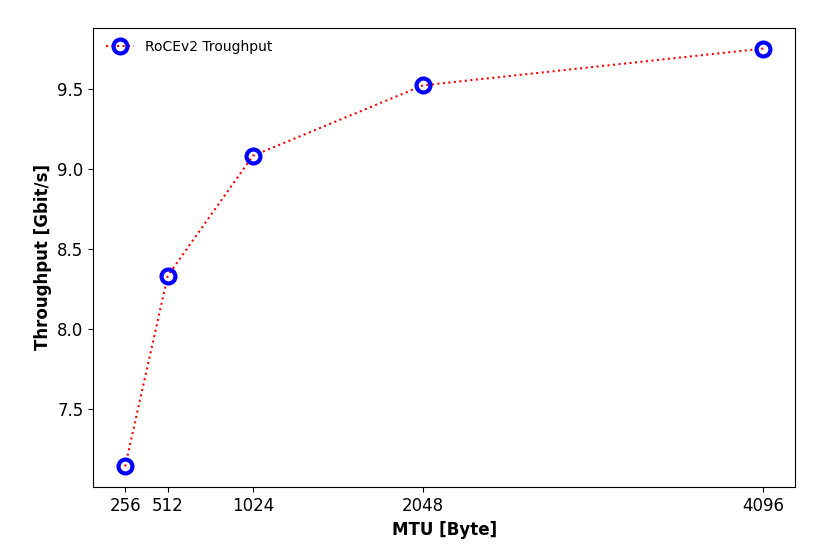}
\caption{Throughput as a function of MTU size for RoCEv2 communication between the KCU105 and a server equipped with a ConnectX-5 NIC.}
\label{roce-throughput}
\end{figure}

For the second test, the target system was upgraded to a server equipped with a Mellanox/NVIDIA ConnectX-5 NIC\cite{connectx5}. The receiver setup again utilized the Pyverbs script, but this time in conjunction with the hardware-based RDMA capabilities of the NIC. The network link between the FPGA and the server was again a 10 Gb/s Ethernet connection.
In this configuration, the throughput reached the limits of the 10 Gb/s link, with the system achieving a sustained data transfer rate of approximately 9.7 Gb/s. This near-line rate performance demonstrated the ability of the RoCEv2 firmware to maximize bandwidth usage when communicating with an RDMA-capable NIC. Fig.~\ref{roce-throughput} shows the throughput achieved with this setup for different MTU sizes. As expected, similarly for the Soft-RoCE test, larger MTU sizes result in higher throughput. 

\section{Experimental testing}

\subsection{Experimental Setup}

\begin{figure}[t]
\centering
\includegraphics[width=3.3in]{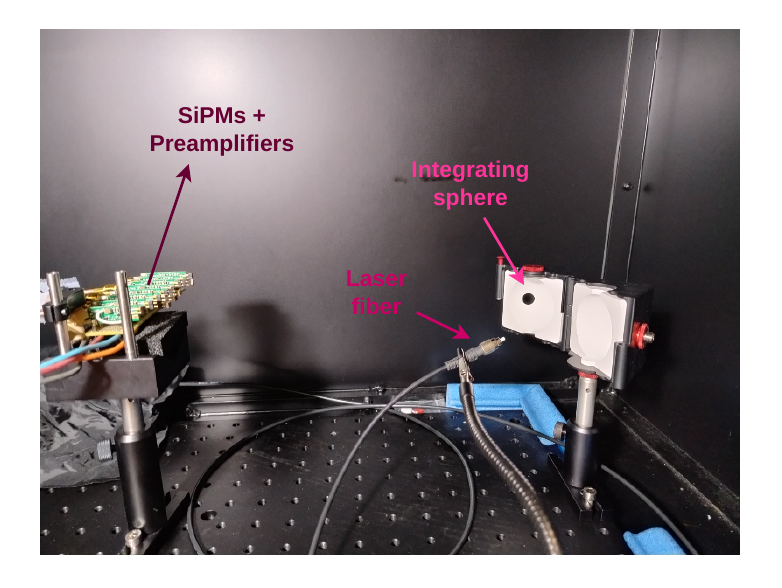}
\caption{The inside of the dark box.}
\label{bb_setup}
\end{figure}

\begin{figure}[t]
\centering
\includegraphics[width=2.5in]{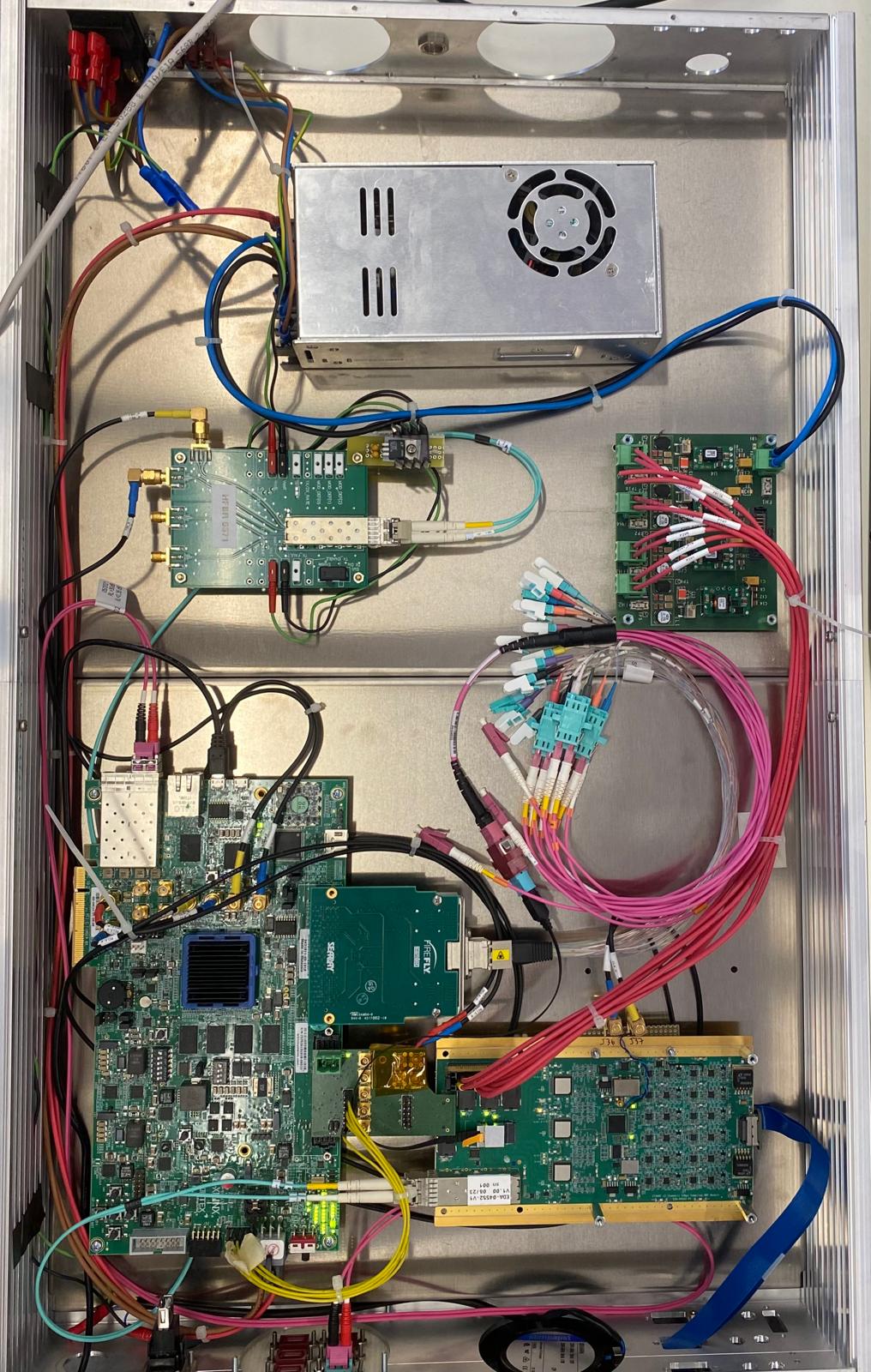}
\caption{The full FE-BE readout system.}
\label{box}
\end{figure}

To evaluate the performance of the FE digitizer board in a realistic scenario, the board was integrated into a test setup simulating operational conditions. A dark, light-tight, box was used to house the core components of the system, visible in Fig.~\ref{bb_setup}: a fiber optic cable, connected to a laser source, delivered light pulses to a SiPM detector \cite{sipm} with similar specifications to those foreseen for use in LST AdvCam. In order for the light to be diffused inside the box and hit every SiPM, the inside of a Thorlabs's integrating sphere\cite{thorlabs} was used. A laser driver \cite{laser} controlled the emission of light pulses, allowing the user to set the timing and intensity characteristics. The driver also provided a 'trigger out' signal, synchronized with the light pulses, to initiate the data acquisition process. To amplify and shape these signals, a custom low-noise preamplifier was employed. The output of the preamplifier was routed outside the box using a coaxial cable, this allows for an easy connection to either the FADC board or to an oscilloscope. Fig.~\ref{box} shows the full FE-BE system used for the test, which has been housed into a 2U rack case to streamline connections and enhance portability. 
The aim of the test was to obtain the typical multi-photoelectron distribution or more commonly called the "finger plot", enabling the discrimination of different photon counts.

\begin{figure}[t]
\centering
\includegraphics[width=3.3in]{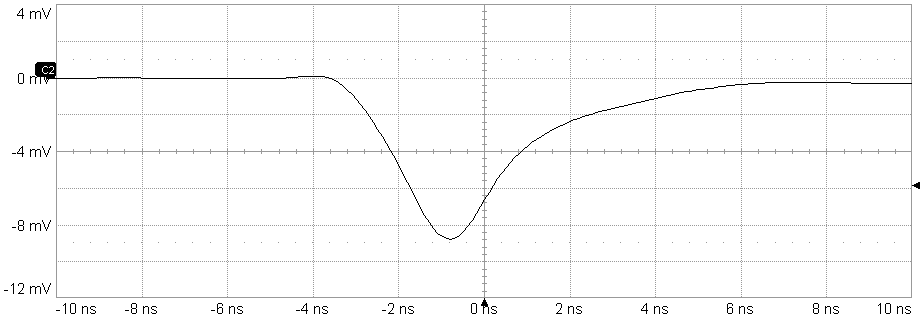}
\caption{A typical SiPM-preamplifier pulse to be acquired by the FADC board.}
\label{typ_pulse}
\end{figure}

Fig.~\ref{typ_pulse} shows a typical waveform from the SiPM preamplifier: fast unipolar pulses with rise times of about 2 ns, fall times around 4 ns, and a bandwidth requirement of up to 200 MHz. This is well within the specifications of the FADC board, whose low-pass anti-aliasing filter is set at 500 MHz.

\subsection{Test Results}
To benchmark the digitizer board's performance, the SiPM output was acquired both with an oscilloscope (12-bit, 20 Gsps) \cite{oscilloscope} and with the custom FADC board. The oscilloscope was configured to measure the area under the curve of each pulse, providing a direct measure of the integrated charge. 

To evaluate the FADC's performance as part of the SiPM readout chain, an offline analysis has then been carried out on the acquired data using Python, where the digitized waveforms were processed to calculate the area under the curve for each pulse. To further enhance the precision of the calculation, given the relatively short duration of the pulses (typically a few nanoseconds) and therefore the limited number of data points captured by the digitizer, a cubic spline interpolation was found to provide a more accurate estimation of the pulse area and therefore applied to the digitized waveforms.

\begin{figure}[t]
\centering
\includegraphics[width=3.2in]{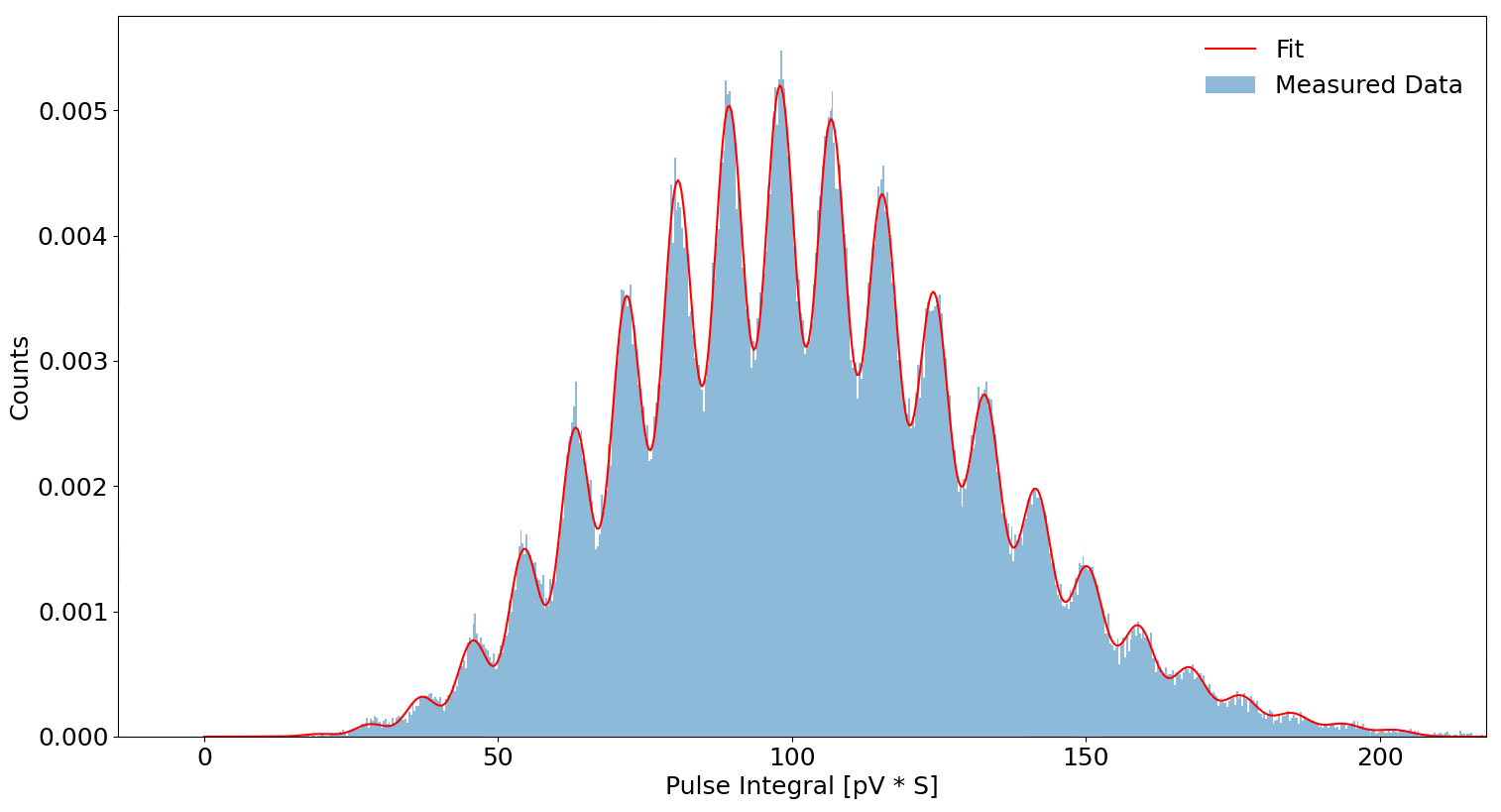}
\caption{Finger plot obtained with the oscilloscope. The red curve illustrates the fit using a generalized Poisson PDF}
\label{finger_osc}
\end{figure}

\begin{figure}[t]
\centering
\includegraphics[width=3.0in]{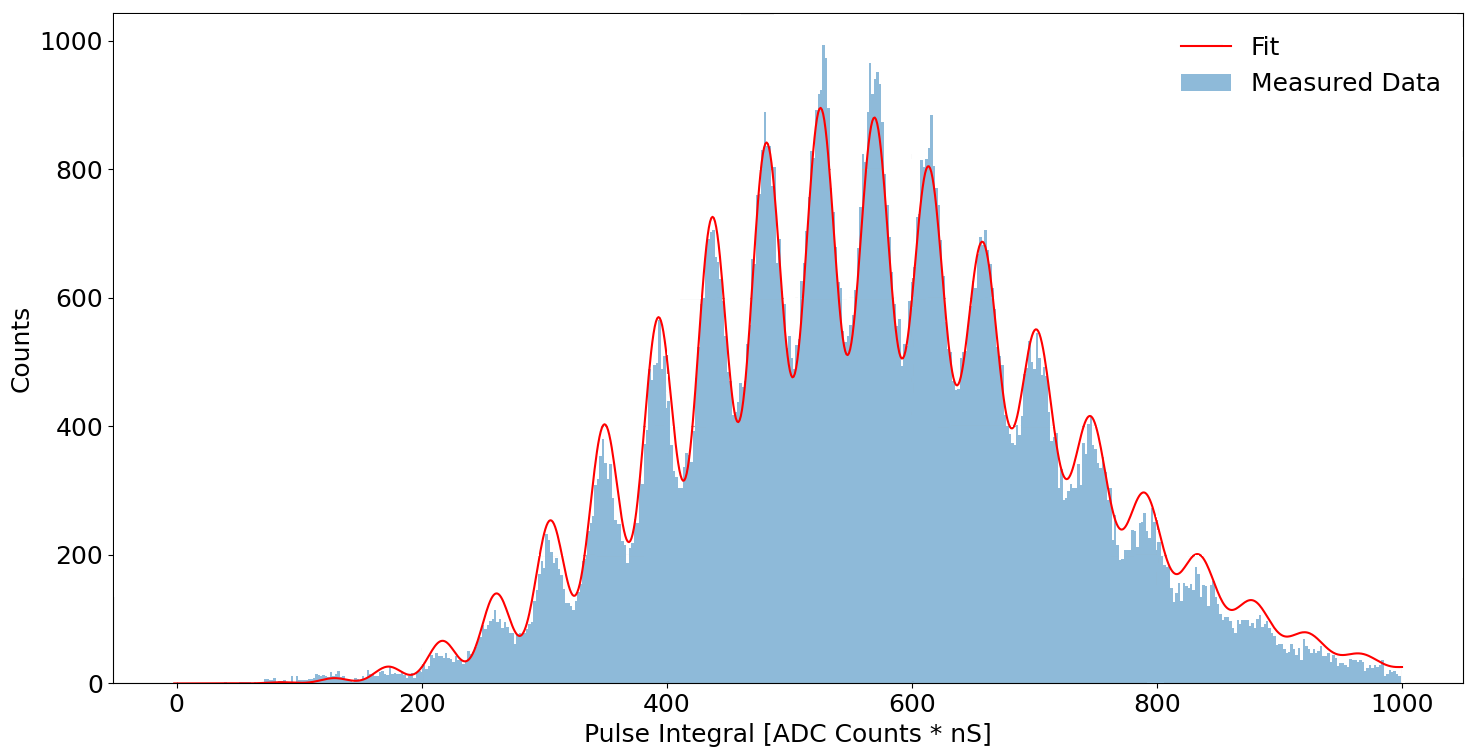}
\caption{Finger plot obtained with the FADC board. The red curve illustrates the fit using a generalized Poisson PDF}
\label{finger_fadc}
\end{figure}

The SNR was adopted as the key performance indicator to assess and compare the quality of the measurements obtained from the oscilloscope and the FADC board. To extract the SNR, the evaluation procedure involved fitting a histogram of the pulse integrals with a generalized Poisson Probability Density Function (PDF), as described in \cite{luca} and \cite{poisson}. The fitted histograms are shown in Fig.~\ref{finger_osc} for the oscilloscope and Fig.~\ref{finger_fadc} for the FADC board. This fitting process yielded several parameters of interest, including the gain, defined as the incremental increase in the pulse integral per detected photoelectron, corresponding to the separation between adjacent multi-photoelectron peaks, and the standard deviations associated with the electronic noise ($\sigma_e$) and sensor-related fluctuations ($\sigma_s$). The SNR was then computed with the following:
\begin{equation}
    \text{SNR}(N_{pe}) = \frac{\text{G}}{\sqrt{\sigma^2_e + N_{pe}\sigma^2_s}}
\end{equation}
For reference, the SNR for 1 p.e. is chosen for the comparison. Using the fit parameters from Tab.~\ref{finger_param}, the SNR was found to be 4.5 for the oscilloscope and 4.0 for the custom digitizer board, with the difference likely attributable to the superior specifications of the oscilloscope, such as its higher resolution and sampling rate. 

The close agreement confirms that the FADC board performs comparably in resolving single-photon events, validating its suitability for precise photon-counting applications.

\begin{table}[]
\caption{Main parameters of the multi-photoelectron response for both the oscilloscope and the FADC board}
\label{finger_param}
\centering
\begin{tabular}{c|c|c|c|}
\cline{2-4}
                                                                                                                                                & \cellcolor[HTML]{EFEFEF}Gain  & \cellcolor[HTML]{EFEFEF}$\sigma_e$ & \cellcolor[HTML]{EFEFEF}$\sigma_s$ \\ \hline
\multicolumn{1}{|c|}{\cellcolor[HTML]{EFEFEF}{\color[HTML]{333333} \begin{tabular}[c]{@{}c@{}}Oscilloscope\\ {[}pV*S{]}\end{tabular}}}          & 32.11 $\pm$ 0.02 & 7.03 $\pm$ 0.08  & 1.18 $\pm$ 0.06    \\ \hline
\multicolumn{1}{|c|}{\cellcolor[HTML]{EFEFEF}{\color[HTML]{333333} \begin{tabular}[c]{@{}c@{}}FADC board\\ {[}ADC Counts*nS{]}\end{tabular}}} & 44.06 $\pm$ 0.05                 & 10.6 $\pm$ 0.5   & 2.8 $\pm$ 0.1      \\ \hline
\end{tabular}
\end{table}

\section{Conclusion and Outlook}
This paper presents the design and development of a novel digital readout electronics system for the R\&D of CTAO's LST Advanced Camera. The system utilizes a custom-designed FADC board for analog-to-digital conversion and a backend board for data processing, trigger generation, and RDMA-based data transfer. The FADC board incorporates high-speed ADCs, a JESD204C interface, and a versatile pre-amplification stage, while the backend board features an FPGA implementation of RDMA technology for efficient data transfer to event-building servers.

The development and implementation of the RoCEv2 core using Bluespec SystemVerilog has enabled the creation of a highly flexible and efficient Target Channel Adapter for specialized applications in environments such as astrophysics experiments. By leveraging the advantages of RDMA over Ethernet, this core ensures high-throughput data transfers while bypassing the receiver's CPU. The removal of RDMA READ operations further optimize the design for cost-sensitive and resource-constrained environments. Additionally, the use of RC mode guarantees reliable data transfer, which is critical for event-building processes in data acquisition systems. The resulting design, which can be scaled for a wide range of Ethernet speeds and FPGA targets, represents a robust solution for custom electronics in physics experiments, particularly in harsh operational conditions.

RoCEv2 standard addresses also the management of network congestion with different algorithms. DCQCN (Data Center Quantized Congestion Notification) is a congestion control algorithm to manage network congestion in data centers \cite{dcqcn}. It combines Explicit Congestion Notification (ECN) with rate-based flow control to prevent packet loss and ensure efficient, high-throughput communication. DCQCN operates by detecting congestion via ECN-marked packets and adjusting the transmission rate dynamically to avoid overwhelming the network switches. This is critical for maintaining the low-latency, lossless nature of RDMA while operating the network as a funnel from backend electronics to DAQ servers. Development is under way to enhance the RDMA core and the SURF MAC core to react to ECN packets according to DCQCN algorithm. A full setup with a DCQCN enabled network switch is in place to qualify the new core implementation within an AdvCam hardware mock-up.

Extensive simulations and tests, both hardware and software, were conducted to ensure the performance and reliability of the system, including signal and power integrity analysis, power consumption measurements and firmware qualifications. The results demonstrate the system's ability to handle high-speed data transmission, and potentially meet the constraints imposed by the telescope's environment.

As mentioned in Sec.~\ref{sec:introduction}, a future development aims to consolidate the functionality of both boards into a single, integrated unit. This will involve incorporating the FPGA directly onto the FADC board, enabling it to handle both the JESD204C data acquisition and the RoCEv2-based data transmission at the FE stage. This single-board solution offers several key advantages:
\begin{itemize}
    \item \textbf{Compactness}: Dramatically reduces system complexity and size. It allows for a direct connection from the SiPMs to the server, with no intervening custom electronics, minimizing potential points of failure and simplifying the integration process.
    \item \textbf{Cost-effectiveness}: Developing a custom electronic board requires a substantial investment of time and financial resources. By eliminating the need for a separate BE system, this approach completely avoids those development costs.
    \item \textbf{Simplified integration}: A single board streamlines integration into the LST Advanced Camera, requiring only off-the-shelf, ECN-compliant Ethernet switches for a complete DAQ chain.
\end{itemize}
To support the increased power consumption of the integrated board, a dedicated cooling structure will be required. A suitable thermal solution is currently being studied to ensure reliable operation in the environmental conditions of the LST Advanced Camera.

This integrated design has the potential to represent a highly efficient, compact, and cost-effective solution for high-speed data acquisition in challenging environments like the LST Advanced Camera and it paves the way for streamlined integration of advanced data transmission technologies in future physics experiments.

\bibliographystyle{IEEEtran}
\bibliography{ref}

\ifCLASSOPTIONcaptionsoff
  \newpage
\fi



%



\end{document}